\documentclass[preprint,12pt]{article}
\usepackage{amssymb}
\usepackage{graphicx}
\usepackage{multirow}
\usepackage{caption}
\usepackage{subfig}
\begin{document}
%%%%%%%%%%%%%%%%%%%%%%%%%%%%%
%\title[]
\title
{Violation of Chandrasekhar Mass Limit: The Exciting Potential of Strongly Magnetized White Dwarfs }
%\smallskip\smallskip
\author{Upasana Das, Banibrata Mukhopadhyay\\ \\
%}
%\author{ 
Department of Physics, Indian Institute of
  Science, Bangalore 560012, India\\
upasana@physics.iisc.ernet.in, bm@physics.iisc.ernet.in}
%\ead{\mailto{upasana@physics.iisc.ernet.in, bm@physics.iisc.ernet.in}}
\maketitle
\begin{center}
Essay written for the Gravity Research Foundation \\ 
2012 Awards for Essays on Gravitation
\end{center}

\vskip1.0cm
\begin{abstract}
We consider a relativistic, degenerate, electron gas under the influence of a strong magnetic field, which describes magnetized white dwarfs. Landau quantization changes the density of states available to the electrons, thus modifying the underlying equation of state. In the presence of very strong magnetic fields a maximum of either one, two or three Landau level(s) is/are occupied. We obtain the mass-radius relations for such white dwarfs and their detailed investigation leads us to propose the existence of white dwarfs having a mass $\sim 2.3M_{\odot}$, which overwhelmingly exceeds the Chandrasekhar mass limit.
\end{abstract}

\newpage

\noindent \textbf{Introduction}
\vskip0.5cm

\noindent Chandrasekhar first showed the maximum possible mass for a white dwarf (WD) to be $\sim 1.44M_{\odot}$ \cite{chandra}. Later, several magnetized WDs have been discovered with surface fields of $10^{5}-10^{9}$ G \cite{kemp}, \cite{putney}, \cite{schmidt}, \cite{reimers}, \cite{jordan}. Anticipating that the central field, which cannot be probed directly, will be much stronger than that on the surface, Ostriker and Hartwick \cite{ost} constructed models of WDs with $B \sim 10^{12}$ G at the center but with a much smaller field at the surface. Recent observations of peculiar Type Ia supernovae - SN2006gz, SN2007if, SN2009dc - seem to suggest super-Chandrasekhar mass WDs as their most likely progenitors \cite{scalzo}. These WDs, found in binary systems, stabilize by rotating and accreting  matter \cite{kato}. In this essay we propose a mechanism by which the WDs exceed the Chandrasekhar mass limit in presence of strong magnetic fields at their centers. In a simple theoretical framework existence of such stars has been reported recently \cite{kundu}.

Here we consider strongly magnetized WDs having interior magnetic field $\gtrsim 10^{15}$ G. These Landau quantized electronic systems have at the most one, two or three occupied Landau level(s) (LL(s)). We obtain an exciting possibility of a mass of WDs $\sim 2.3M_{\odot}$.

\vskip1.0cm

\noindent \textbf{Equation of state and solution procedure}
\vskip0.5cm

\noindent \textit{Basic equations}
\vskip0.5cm

\noindent The Landau quantized energy states of a free electron in a uniform, static, magnetic field are given by \cite{lai}  

\begin{equation}
E_{\nu,\,p_{z}} = [p_{z}^{2}c^{2} + m_{e}^{2}c^{4}(1 + \nu \frac{2B}{B_{c}})]^{1/2},
\label{dirac}
\end{equation}
where $\nu$ denotes the LL, given by
\begin{equation}
\nu = j + \frac{1}{2} + \sigma , 
\end{equation}
\noindent when 
$j$ being the principal quantum number, $\sigma = \pm \frac{1}{2}$, $p_{z}$ the momentum of the electron along the $z$-axis, $B$ the magnetic field, $B_{c}$ a critical magnetic field defined by 
\begin{equation}
B_{c} = m_{e}^{2}c^{3}/\hbar e = 4.414 \times 10^{13} \, G,
\end{equation} 
where $e$ is the charge of the electron, $c$ the speed of light, $m_{e}$ the rest-mass of the electron, $\hbar$ the Planck's constant. Our interest is to study the effect of $B > B_{c}$ on the relativistic, degenerate electron gas.

%Energy states of a free electron in a uniform magnetic field is Landau quantized. We define a critical magnetic field strength \cite{lai}
%\begin{equation}
%B_{c} = m_{e}^{2}c^{3}/\hbar e = 4.414 \times 10^{13} \, G,
%\end{equation} 
%where $e$ is the charge of the electron, $c$ the speed of light, $m_{e}$ the rest-mass of the electron and $\hbar$ the Planck's constant. Now the electrons become relativistic either due to high density, when their mean Fermi energy exceeds $m_{e}c^{2}$, or when due to strong magnetic field their cyclotron energy exceeds $m_{e}c^{2}$. 

%Thus in order to study the effect of $B \gtrsim B_{c}$ on the equation of state of a relativistic, degenerate electron gas we solve the Dirac equation in a uniform, static magnetic field directed along the $z$-axis. The energy eigenstates are \cite{lai}
%\begin{equation}
%E_{\nu,\,p_{z}} = [p_{z}^{2}c^{2} + m_{e}^{2}c^{4}(1 + \nu \frac{2B}{B_{c}})]^{1/2},
%\label{dirac}
%\end{equation}
%where quantum number $\nu$ denotes the Landau level and is given by
%\begin{equation}
%\nu = j + \frac{1}{2} + \sigma , 
%\end{equation}
%$j$ being the principal quantum number of the Landau level ($j$ = 0, 1, 2,...), $\sigma = \pm \frac{1}{2}$, the spin of the electron, $p_{z}$ the momentum of the electron along the $z$-axis and $B$ the magnetic field. 

%\noindent The strong magnetic field modifies the density of states as follows
%\begin{equation}
%\frac{2}{h^{3}}\int d^{3}p    \longrightarrow   \sum_{\nu} \frac{2eB}{h^{2}c}\,g_{\nu} \int dp_{z},
%\end{equation}
%where $g_{\nu} = 1$ for $\nu = 0$ and $g_{\nu} = 2$ for $\nu \geq 1$. 

\noindent The Fermi energy of electrons in units of $m_{e}c^{2}$ for a given $\nu$ is given by 
\begin{equation}
\epsilon_{F}^{2} = x_{F}(\nu)^{2} + (1 + 2\nu \frac{B}{B_{c}}),
\label{fermi}
\end{equation}
where $x_{F}(\nu)$ is the Fermi momentum in units $m_{e}c$.

\noindent As $x_{F}(\nu)^{2} \geq 0$, the maximum number of occupied LLs 
\begin{equation}
\nu_{m} = \left(\frac{\epsilon_{Fmax}^{2} - 1}{2B_{D}}\right)_{\rm nearest ~ lowest ~ integer},
\label{max}
\end{equation}
where $B_{D} = B/B_{c}$, $\epsilon_{Fmax}$ the dimensionless maximum Fermi energy of the system.

\noindent Following \cite{lai} we write the electron number density 
\begin{equation}
n_{e} = \frac{2B_{D}}{(2\pi)^{2} \lambda_{e}^{3}} \sum_{\nu=0}^{\nu_{m}} g_{\nu}x_{F}(\nu),
\label{ne}
\end{equation}
where the Compton wavelength of the electron $\lambda_{e} = \hbar/m_{e}c$; the matter density
\begin{equation}
\rho = \mu_{e}m_{H}n_{e},
\label{mat}
\end{equation}
where $\mu_{e}$ is the mean molecular weight per electron (which we choose to be $2$) and $m_{H}$ the mass of hydrogen atom; the electron degeneracy pressure 

%\noindent The electron energy density
%\begin{eqnarray}  
%\varepsilon_{e} & = & \frac{2B_{D}}{(2\pi)^{2}\lambda_{e}^{3}}\sum_{\nu=0}^{\nu_{m}} g_{\nu} \int\limits_{0}^{x_{e}(\nu)} E_{\nu,\,p_{z}}d \left (\frac{p_{z}}{m_{e}c} \right) \nonumber \\ 
%& = & \frac{2B_{D}}{(2\pi)^{2}\lambda_{e}^{3}}m_{e}c^{2}\sum_{\nu=0}^{\nu_{m}} g_{\nu} (1 + 2\nu B_{D}) \bold{\psi} \left (\frac{x_{e}(\nu)}{(1 + 2\nu B_{D})^{1/2}} \right), 
%\label{endensity}
%\end{eqnarray}
%where
%\begin{equation}
%\bold{\psi}(x) = \int\limits_{0}^{x} (1 + y^{2})^{1/2}dy = \frac{1}{2}x \sqrt{1 + x^{2}} + \frac{1}{2}\ln(x + \sqrt{1 + x^{2}}).
%\end{equation}

\begin{eqnarray}
P =  \frac{2B_{D}}{(2\pi)^{2}\lambda_{e}^{3}}m_{e}c^{2}\sum_{\nu=0}^{\nu_{m}} g_{\nu} (1 + 2\nu B_{D}) \bold{\eta} \left (\frac{x_{F}(\nu)}{(1 + 2\nu B_{D})^
{1/2}} \right), 
\label{pressure}
\end{eqnarray}
where
\begin{equation}
\bold{\eta}(y) = \frac{1}{2}y\sqrt{1 + y^{2}} - \frac{1}{2}\ln(y+ \sqrt{1 + y^{2}}).
\end{equation}

\vskip0.5cm
\noindent \textit{Procedure}
\vskip0.5cm

\noindent From equation (\ref{max}) we see that if $B_{D} \gg 1$, then $\nu_{m}$ is small, implying electrons are restricted to the lower LLs. Since we are interested in this regime, depending on the specific values of $B_{D}$ and $\epsilon_{Fmax}$, we have one-level ($0 \leq \nu_{m} < 1$), two-level ($1 \leq \nu_{m} < 2$) and three-level ($2 \leq \nu_{m} < 3$) systems. By eliminating $x_{F}(\nu)$ numerically from equations (\ref{mat}) and (\ref{pressure}), we obtain the $P-\rho$ relation which is the equation of state (EOS). Figure 1 shows EOSs for the different cases given in Table 1.

\begin{table}[h]
\vskip0.2cm
{\centerline{\large Table 1}}
{\centerline{ Parameters for the equations of state in Figure 1. }}
{\centerline{}}
\begin{center}
\begin{tabular}{|c|c|c|c|}

\hline
$\epsilon_{Fmax}$ & $\nu_{m}$ &  $B_{D}$ & $B$ in units of $10^{15}$ G\\
\hline

%\multirow{3}{*}{2} & 1 & 1.5 & 0.066 \\
& 1 & 1.5 & 0.066 \\
2 & 2 & 0.75 &  0.033 \\
& 3 & 0.5 &  0.022 \\ \hline
%\multirow{3}{*}{20} & 1 & 199.5 &  8.81 \\
& 1 & 199.5 &  8.81 \\
20 & 2 & 99.75 &  4.40 \\
& 3 & 66.5 & 2.94  \\ \hline
%\multirow{3}{*}{200} & 1 & 19999.5 & 882.78 \\
& 1 & 19999.5 & 882.78 \\
200 & 2 & 9999.75 & 441.38 \\
& 3 & 6666.5 &  294.26 \\ \hline

\end{tabular}

\end{center}
\end{table}

%we fix $B_{D}$, such that it can only take values 1, 2 or 3, which we call as one-level, two-level and three-level systems respectively. By one-level system we mean where only the ground Landau level, $\nu = 0$, is occupied, two-level means where both the ground and the first ($\nu = 1$) Landau levels are occupied and three-level means where the ground, first and second ($\nu = 2$) Landau levels are occupied. Thus once we fix $\nu_{m}$, we obtain a fixed $B_{D}$ on supplying a desired $\epsilon_{Fmax}$ from equation (\ref{max}). We choose $E_{Fmax} = 2\, m_{e}c^{2}$, $20\, m_{e}c^{2}$ and $200\, m_{e}c^{2}$ and for each we study the $\nu_{m}=1,2$ and $3$ systems, which are listed in Table 1. For each of these cases we obtain the equation of state by simultaneously solving equations (\ref{ne}), (\ref{mat}) and (\ref{pressure}), numerically from $E_{F} =  m_{e}c^{2}$ to $E_{F} = E_{Fmax}$, when each value of $E_{F}$ gives one point in the $P-\rho$ plot. Figure 1 shows the equation of states for the different cases given in Table 1.

%If we are to construct the model of a magnetized white dwarf, we require to solve the following differential equation which comes from the condition of hydrostatic equilibrium \cite{arc}:

The hydrostatic equilibrium of the WD in presence of constant magnetic field is described by \cite{arc}
\begin{equation}
\frac{1}{r^{2}}\frac{d}{dr}\left(\frac{r^{2}}{\rho}\frac{dP}{dr} \right) = -4\pi G \rho,
\label{diff}
\end{equation}
\noindent with the boundary conditions:

\begin{equation}
\rho(r = 0) = \rho_{c} 
\end{equation}
and
\begin{equation}
\left (\frac{d\rho}{dr} \right)_{r=0} = 0,
\end{equation}
\noindent where $\rho_{c}$ is the central density of the WD.
%Since the pressure cannot be expressed as an analytical function of density, unlike that of Chandrasekhar's work \cite{chandra}, we fit the equation of state with the following polytropic relation:
%\begin{equation}
% P = K\rho^{\Gamma},
%\label{poly}
%\end{equation}
%with different values of the adiabatic index $\Gamma$ in different density ranges ($K$ being a dimensional constant). Thus the actual equation of state is reconstructed using multiple polytropic equations of state. This way we can determine the effect of magnetic field on the adiabatic index of the matter. One such fit is shown in Figure 1(d) and the parameters $K$ and $\Gamma$ are stated in Table 2. 

The radius $R$ of the WD is obtained by solving equation (\ref{diff}) and is the value of $r$ where density goes to zero. The mass $M$ of the WD, which is approximated to be spherical, is obtained by integrating the following equation:
\begin{equation}
\frac{dM}{dr} = 4\pi r^{2}\rho.
\label{mass}
\end{equation}

\captionsetup[subfigure]{position=top}
\begin{figure}[h]
  \begin{center}
    \begin{tabular}{ll}
 \subfloat[]{\includegraphics[scale=1.]{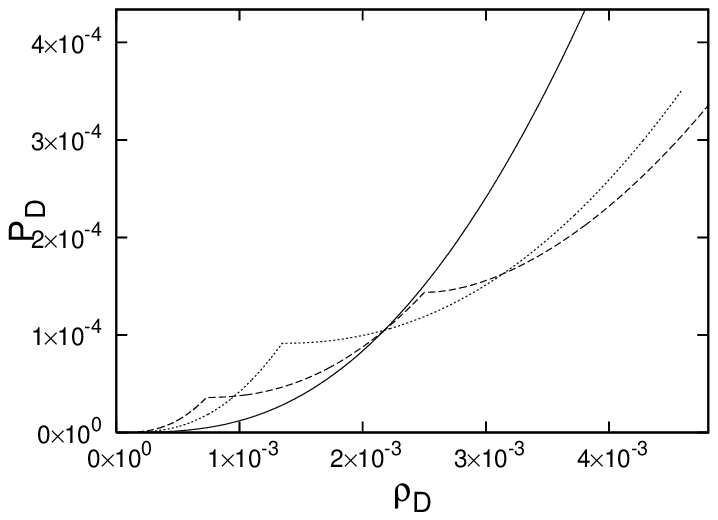}}&
   \subfloat[]{\includegraphics[scale=1.]{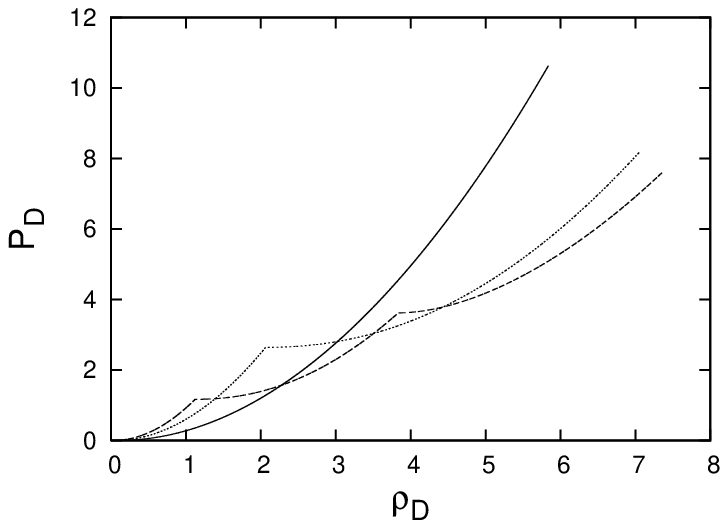}} \\ \\ \\
    \subfloat[]{\includegraphics[scale=1.]{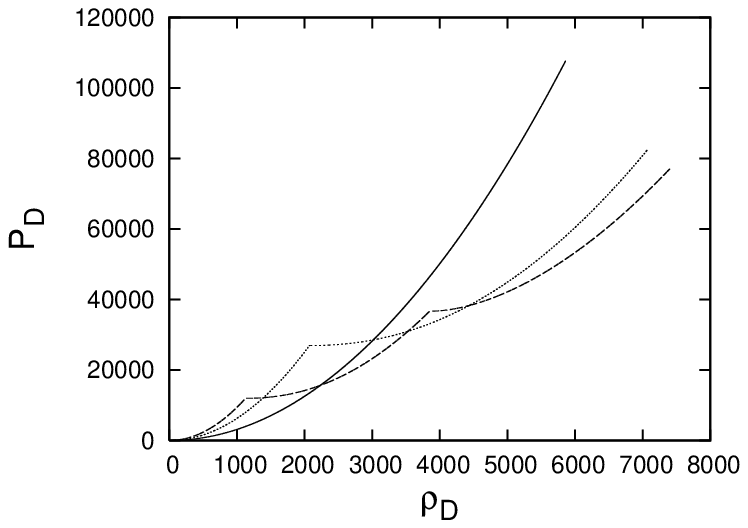}}&
 \subfloat[]{\includegraphics[scale=1.]{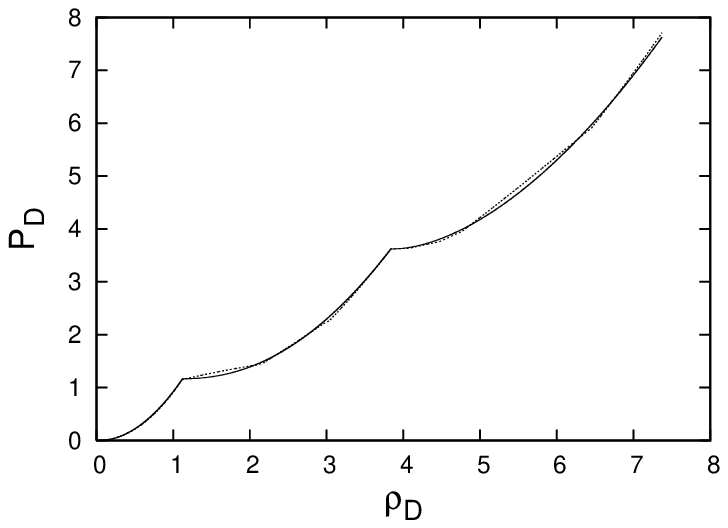}} \\
      \end{tabular}
    \caption{Equations of state in a strong magnetic field (given in Table 1) for (a) $\epsilon_{Fmax} = 2$, (b) $\epsilon_{Fmax} = 20$, (c) $\epsilon_{Fmax} = 200$. In all three cases the solid, dotted and dashed lines indicate one-level, two-level and three-level systems respectively. In (d) the solid line is same as the dashed line in (b), but fitted with the dotted line by analytical formalism (see text for details). Here $P_{D}$ is the pressure in units of $2.668 \times 10^{27}$ erg/cc and $\rho_{D}$ is the density in units of $2 \times 10^{9}$ gm/cc.}
    \label{eos}
  \end{center}
\end{figure}
%\noindent A plot of $R$ as a function of $M$ gives the mass-radius relation for the magnetized white dwarf, which is shown in Figure 2. 
%\begin{figure}[h]
%\centering
 % \subfloat[1 level]{\label{fig:1level}\includegraphics[scale=0.4, trim=1.5cm 0cm 0cm 0cm, clip = true,angle=270]{1level.ps}} 
 % \subfloat[2 level]{\label{fig:2level}\includegraphics[scale=0.4,angle=270]{2level.ps}}
 % \subfloat[3 level]{\label{fig:3level}\includegraphics[scale=0.4,angle=270]{3level.ps}}
%  \caption{Equation of State}
 % \label{fig 1}
%\end{figure}Table 2

\vskip1.0cm
\noindent \textbf{Constructing white dwarfs}
\vskip0.5cm
\noindent \textit{Equations of State}
\vskip0.5cm

%\noindent Now we come to the discussions of the results obtained. Table 1 gives the magnetic field strengths corresponding to the different systems. We mention here that for a given value of $E_{Fmax}$, the value of $B_{D}$ listed here corresponds to a lower limit. For example, when $E_{Fmax} = 20\,m_{e}c^{2}$, $B_{D}$ with a value of 199.5 just results in a one-level system but, if we choose any $B_{D} >$ 199.5 that would also lead to a one-level system.

Let us consider the case $\epsilon_{Fmax} = 20$. The solid curve in Figure 1(b) is free of any kink (only ground LL occupied). The dotted and dashed curves have one kink (ground and first LLs occupied) and two kinks (ground, first and second LLs occupied) respectively. Any kink represents a transition from a lower to upper LL.

%Let us consider Figure 1(b) which shows the equations of state with $E_{Fmax} = 20\, m_{e}c^{2}$. We notice that the solid curve ($\nu_{m}=1$) is free of any kink, the dotted curve ($\nu_{m}=2$) has one kink and the dashed curve ($\nu_{m}=3$) has two kinks. The kinks appear whenever there is a transition from a lower Landau level to the next and they denote regions of the equation of state where the pressure becomes independent of density. For $\nu_{m}=1$ there re no kinks as all the electrons are in the ground Landau level. For $\nu_{m}=2$ the portion of the equation of state below the kink represents the ground Landau level and the one above represents the first Landau level. As the Fermi energy of the electrons increases, more and more electrons occupy the ground Landau level and both the density and pressure of the system keep increasing. Once the ground level is completely filled, one observes that, on increasing the Fermi energy of the electrons the density increases but, the pressure remains fairly constant for a while, after which the pressure again starts increasing with density. It is as if, the increase in Fermi energy during the transition causes the system to move to a higher Landau level instead of increasing the pressure. This situation seems analogous to that of phase transition in matter (where the temperature remains constant with respect to the input heat energy during the change of phase). 

\vskip0.5cm
\noindent \textit{Mass-radius relations}
\vskip0.5cm

\captionsetup[subfigure]{position=top}
\begin{figure}[h]
  \begin{center}
    \begin{tabular}{ll}
 \subfloat[]{\includegraphics[scale=1.]{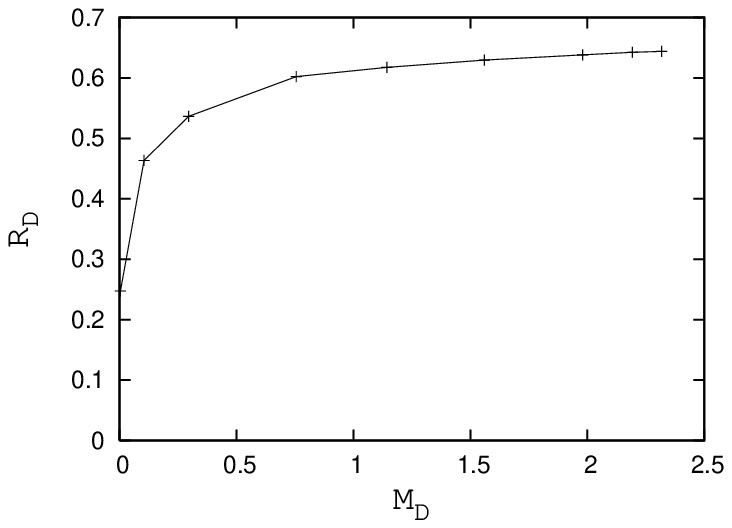}}&
   \subfloat[]{\includegraphics[scale=1.]{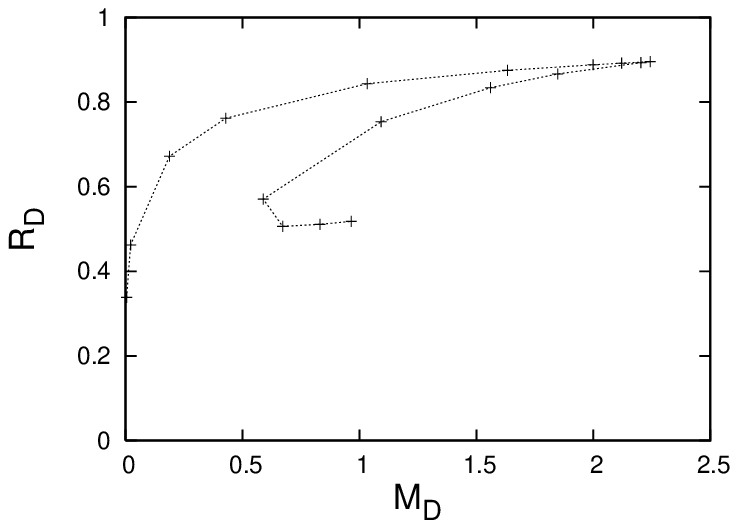}} \\ \\ \\
    \subfloat[]{\includegraphics[scale=1.]{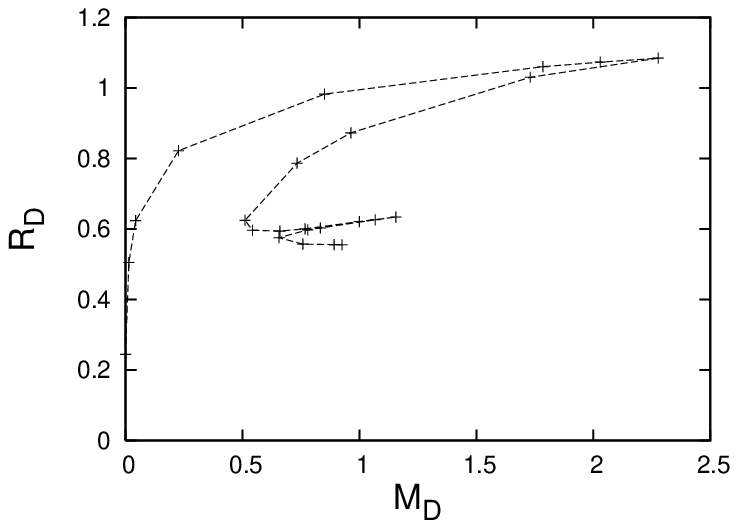}}
           \end{tabular}
    \caption{Mass-radius relations with $\epsilon_{Fmax} = 20$ for (a) one-level system, (b) two-level system, (c) three-level system. Here $M_D$ is the mass of the white dwarf in units of $M_{\odot}$ and $R_D$ is its radius in units of $10^{8}$ cm (the solid, dotted and dashed lines have the same meaning as in Figure 1).}
    \label{mr1}
  \end{center}
\end{figure}

\begin{table}[h]
\vskip0.2cm {\centerline{\large Table 2}}{\begin{center} Parameters for the fitting function for the equation of state shown 
in Figure 1(d).\end{center}}

\begin{center}
\begin{tabular}{|c|c|c|}

\hline
$\rho_{D}$ in units of $2 \times 10^{9}$gm/cc & $\Gamma$ & $K$ in CGS units \\
\hline

0 $-$ 0.096 & 2.9 & 4.055 \\
0.096 $-$ 0.307 & 2.4 & 1.284  \\ 
0.307 $-$ 1.128 & 2.1 & 0.914  \\
1.128 $-$ 2.117 & 0.35 & 1.105 \\
2.117 $-$ 2.956 & 4/3 & 0.522 \\
2.956 $-$ 3.842 & 2.0 & 0.246  \\
3.842 $-$ 4.651 & 0.35 & 2.225  \\
4.651 $-$ 6.116 & 4/3 & 0.496 \\
6.116 $-$ 7.37 & 2.0 & 0.142  \\ \hline

\end{tabular}

\end{center}

\end{table}

\noindent Figure 2 shows the mass-radius relations, where each point in the curves corresponds to a WD with a particular value of $\rho_{c}$ which is supplied as a boundary condition.

%\noindent Next we come to Figure 2, where we show the mass-radius relations for the $\nu_{m}=1,2$ and $3$ systems ((a), (b) and (c) respectively), with $E_{Fmax} = 20\,m_{e}c^{2}$ (results for $E_{Fmax} = 2\, m_{e}c^{2}$ and $200\, m_{e}c^{2}$ also show the same trend). Each point in the $R_{D}-M_{D}$ curve corresponds to a star with a particular value of central density $\rho_{c}$ which is supplied as a boundary condition. ($R_{D}$ and $M_{D}$ are the dimensionless radius and mass of a star respectively, as defined in Figure 2 caption.)

In Figure 2(a) we see that as $\rho_{c}$ increases, both $M_{D}$ and $R_{D}$ increase and then at higher values of $\rho_{c}$, $R_{D}$  becomes nearly independent of $M_{D}$. We note that the most massive WD on this curve has a mass $\sim$ 2.3 $M_{\odot}$, which corresponds to the maximum density point of the solid curve in Figure 1(b). This denotes the density at which the ground LL is completely filled. Thus a WD with this $\rho_{c}$ and a magnetic field strength of $B = 8.81 \times 10^{15}$ G has a mass greater than the Chandrasekhar limit (for details see \cite{das}). 

The turning point in Figure 2(b), after attending the maximum mass $\sim 2.3 M_{\odot}$, corresponds to the kink in the corresponding EOS given in Figure 1(b). During the transition from ground to first LL, the radius and mass both decrease with increasing $\rho_{c}$. Then there is a brief range of densities where the mass increases as the radius decreases and ultimately at very high densities the radius is nearly independent of the mass. 

In Figure 2(c) we see a repetition of the features in Figure 2(b) twice, which corresponds to the two kinks in the corresponding EOS in Figure 1(b). (The maximum mass $\sim 2.3 M_{\odot}$ occurs at that $\rho_{c}$ where the ground LL is completely filled and transition to the first LL is about to start.) 

%In Figure 2(c) we see two turning points which correspond to the two kinks in the corresponding equation of state (dashed curve in Figure 1(b)). The maximum mass (at the first turning point) $\sim 2.3 M_{\odot}$ is reached at a radius of $\sim 1.1 \times 10^{8}$ cm for $\rho_{c} \sim 2.2 \times 10^{9}$ gm/cc. The mass at the second turning point is $\sim 1.2 M_{\odot}$ which has a radius $\sim 6.3 \times 10^{7}$ cm for $\rho_{c} \sim 7.6 \times 10^{9}$ gm/cc. Just after either of the turning points the radius and mass decrease, then briefly the radius decreases as the mass increases and finally the radius becoming nearly independent of mass. The maximum mass $\sim 2.3 M_{\odot}$ occurs at the central density where the ground Landau level is completely filled and transition to the first Landau level is about to start. 

For comparison let us recall the Lane-Emden solution for the classical non-magnetic case, which assumes a polytropic EOS
\begin{equation}
 P = K\rho^{\Gamma},
\label{poly}
\end{equation}
\noindent giving rise to the following relations \cite{arc}:
\begin{equation}
R \propto \rho_{c}^{\frac{\Gamma - 2}{2}}
\label{rvsgamma}
\end{equation}
and
\begin{equation}
M \propto \rho_{c}^{\frac{3\Gamma -4}{2}}.
\label{mvsgamma}
\end{equation}

\noindent This means that if $\Gamma > 2$, $R$ increases with $\rho_{c}$, if $\Gamma = 2$, $R$ is independent of $\rho_{c}$, if $\Gamma > 4/3$, $M$ increases with $\rho_{c}$ and if $\Gamma = 4/3$, $M$ is independent of $\rho_{c}$. This is exactly what is observed in Figure 2, if the EOSs given in Figure 1 are fitted by the polytropic EOS, adopting constant values of $\Gamma$ in different density ranges (see Table 2). For instance in Figure 1(b) (dashed line) and Figure 2(c), at very low densities, $\Gamma=3$ and up to the turning point density (the kink in the EOS) the radius keeps increasing with mass and then becomes nearly independent of mass when $\Gamma \sim 2$. Then $\Gamma$ suddenly drops to a small value $\sim$ 0.35 which marks the onset of the transition from ground LL to first LL. In this region the pressure becomes independent of density, revealing an unstable zone in the EOS. As the density increases further, $\Gamma$ approaches the relativistic value of $4/3$. In this regime we see that the radius decreases slightly as the mass does not change significantly. Next $\Gamma$ takes up a value of 2 and the above phenomena will be repeated.

\captionsetup[subfigure]{position=top}
\begin{figure}[h]
  \begin{center}
    \begin{tabular}{ll}
 \subfloat[]{\includegraphics[scale=0.75]{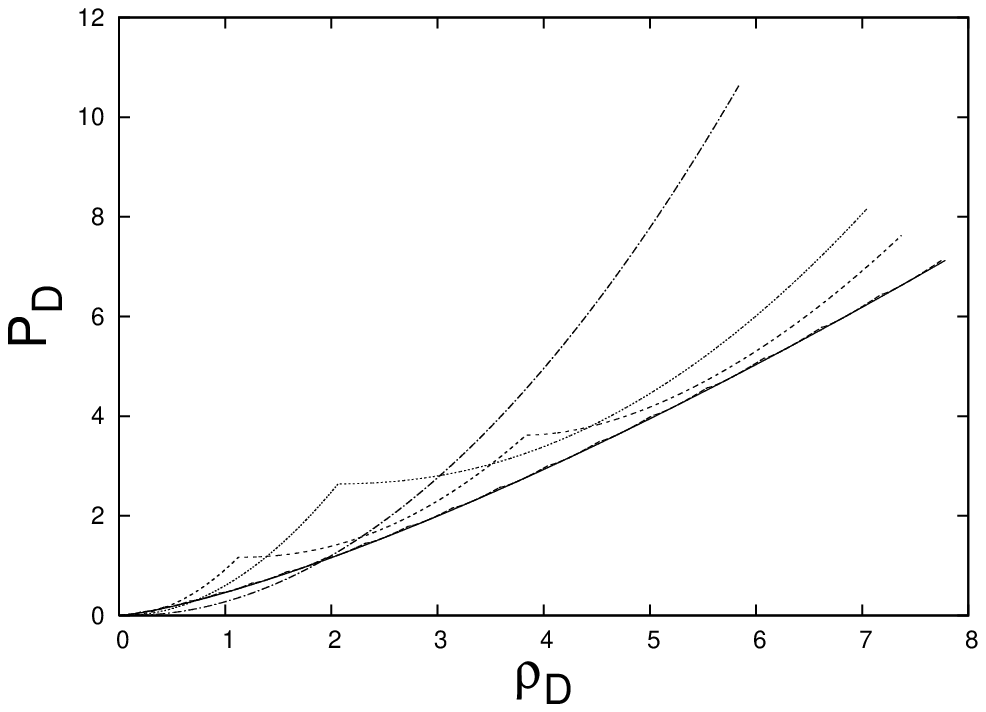}}&
  \subfloat[]{\includegraphics[scale=0.75]{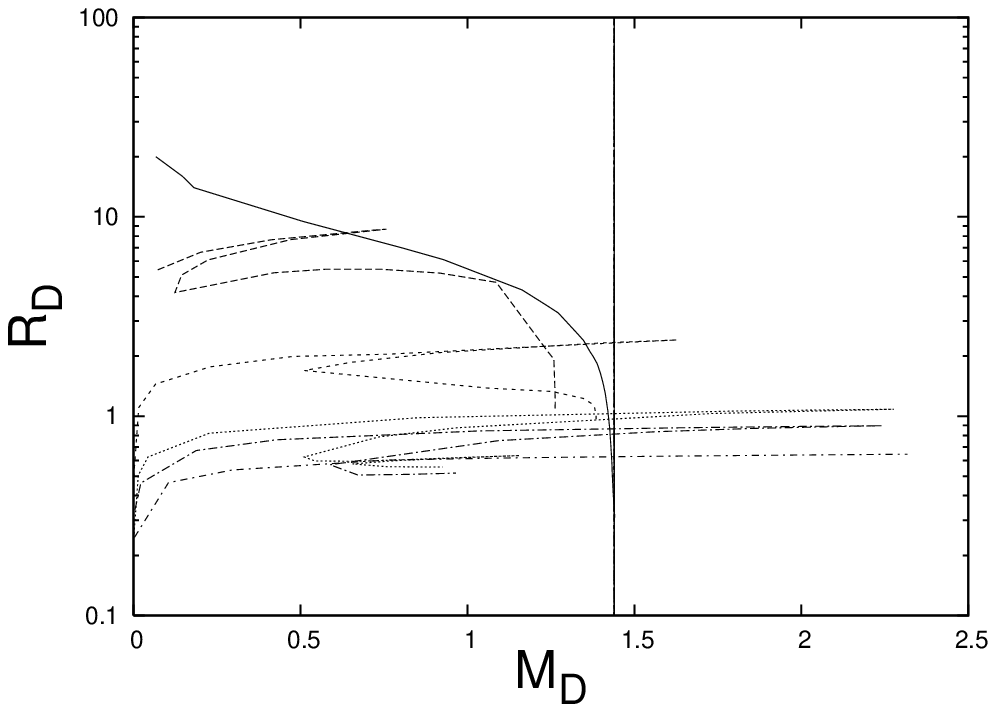}}
           \end{tabular}
    \caption{Comparison with Chandrasekhar's non-magnetic results for $\epsilon_{Fmax} = 20$. (a) Equations of state - the solid line represents Chandrasekhar's equation of state. The dot-dashed, dotted and dashed lines represent the one-level, two-level and three-level systems respectively. The equation of state for $\nu_{m} = 20$ is also shown, which appears as a series of kinks on top of the solid line. (b) Mass-radius relations - the vertical line marks the 1.44$M_{\odot}$ limit and the solid line represents Chandrasekhar's mass-radius relation. From top to bottom the other lines represent $\nu_{m} = 500\,, 20\,, 3\,, 2$ and $1$ respectively (the $y$-axis is in log scale).}
   \label{nonmagnetic}
  \end{center}
\end{figure}

\captionsetup[subfigure]{position=top}
\begin{figure}[h]
  \begin{center}
    \begin{tabular}{ll}
 \subfloat[]{\includegraphics[scale=0.8]{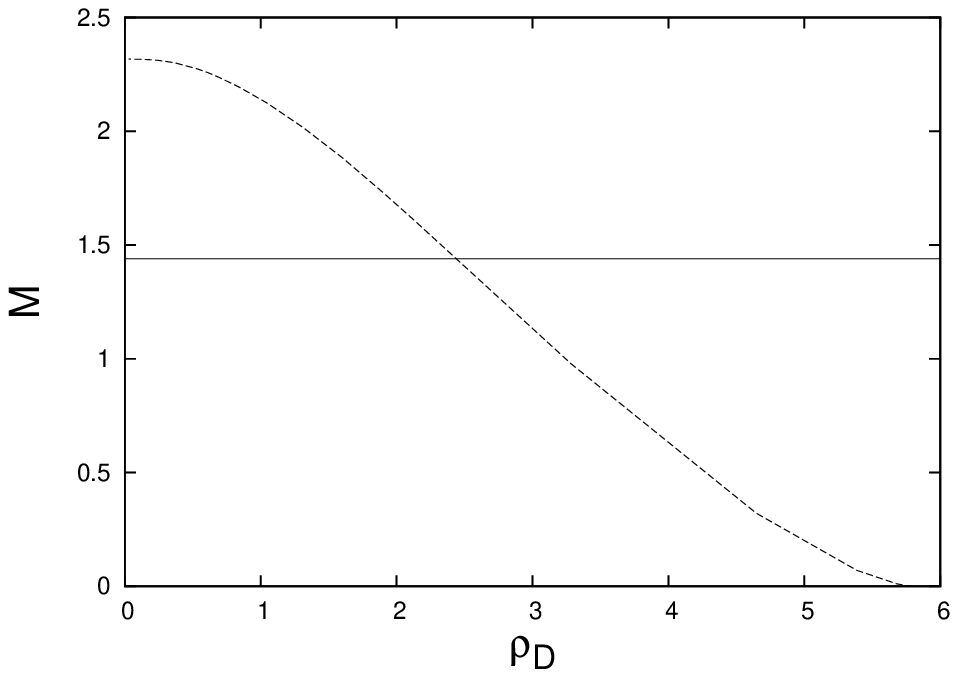}}&
   \subfloat[]{\includegraphics[scale=0.8]{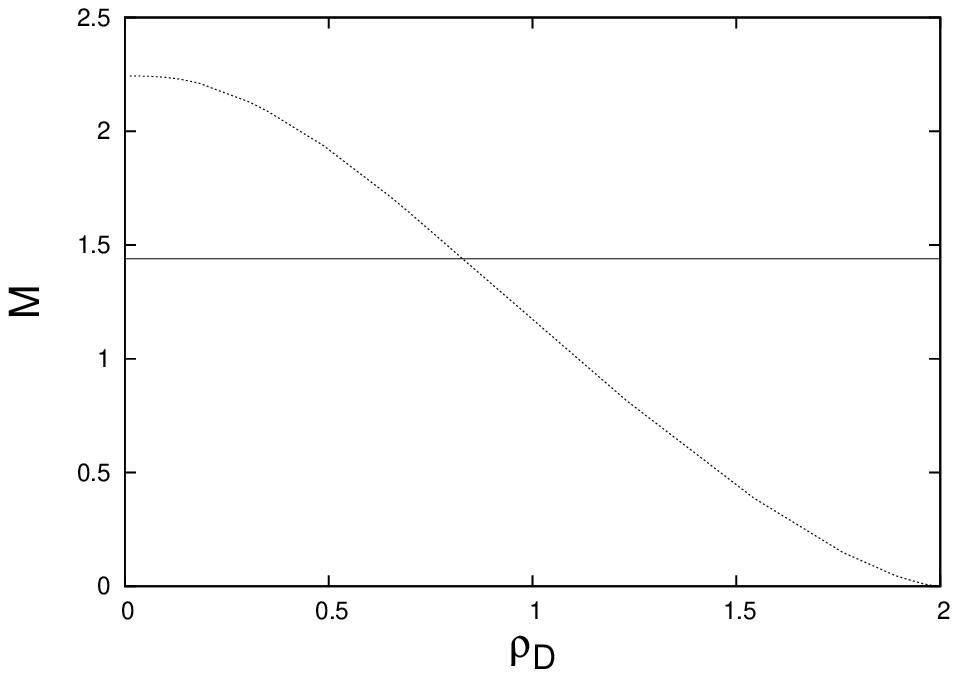}} \\ \\ \\
    \subfloat[]{\includegraphics[scale=0.8]{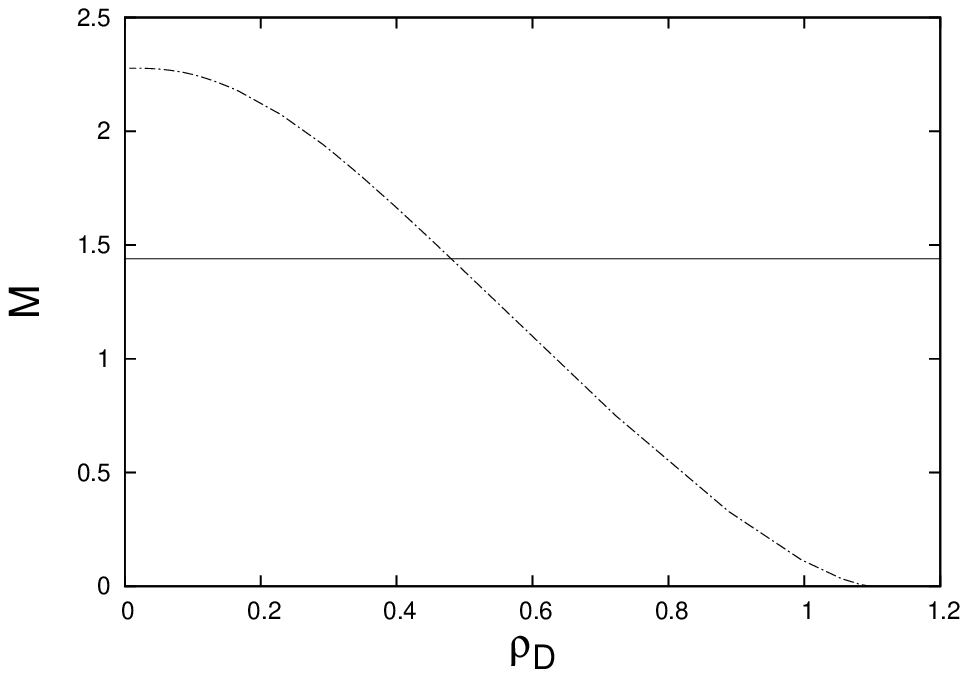}} 
       \end{tabular}
    \caption{Mass as a function of density inside a magnetized white dwarf with $B_{D}$ (a) 199.5, (b) 99.75, (c) 66.5, with $\epsilon_{Fmax} = 20$. The horizontal line indicates the 1.44$M_{\odot}$ limit. See Table 1 for details.}
   \label{mrho}
  \end{center}
\end{figure}

\vskip0.5cm
\noindent \textit{Discussions}
\vskip0.5cm

\noindent In Figure 3 we put together our results along with that of Chandrasekhar. Figure 3(a) clearly indicates that as the magnetic field decreases, the EOS approaches Chandrasekhar's EOS.

Figure 3(b) represents the corresponding mass-radius relations. We observe that as the magnetic field decreases, the mass-radius relation approaches that obtained by Chandrasekhar. We also note that as the magnetic field increases the WDs become more and more compact in size and the probability that they will have masses exceeding the Chandrasekhar limit also increases.

Figure 4 shows the variation of mass as a function of density within a magnetized WD for three different magnetic fields. In all the cases, we note that, by the time the density falls to about half the value of $\rho_{c}$, the mass has already crossed the Chandrasekhar limit as indicated by the horizontal line. Hence the effect of magnetic field is restricted to the high density regime, where the field remains essentially constant even in reality. This justifies our choice of constant field.

%This would be more clear from the description of the variation of magnetic field given by \cite{monica}, \cite{bandyo}, which show that an inhomogeneous magnetic profile can be chosen such that the magnetic field is nearly constant throughout most of the star and then falls off close to the surface (see Figure 5(b) in \cite{monica}). Thus choosing an inhomogeneous magnetic profile will not affect our main finding that, the Chandrasekhar mass limit can be exceeded for high magnetic field strengths. 

\vskip1.0cm
\noindent \textbf{Conclusions}
\vskip0.5cm 

%\noindent We have studied the effect of high magnetic field on the equation of state of purely electron degenerate matter at zero temperature. In the equation of state, we have considered only the electron degeneracy pressure modified by the strong magnetic field. We have focused on those Landau quantized systems in which the maximum number of Landau levels occupied are one, two or three, which we have named to be one-level, two-level and three-level systems respectively. 

\noindent We summarize the important findings of this work as follows: 
\begin{itemize}

\item A transition from the lower to upper LL represents a kink in the EOS. If $\rho_{c}$ lies at the kink appearing at the ground to first LL transition, the WD has the maximum possible mass.

%\item We have found that whenever a lower Landau level is completely filled and the next higher level is to be filled, a kink appears in the equation of state - which is a small region where the pressure becomes nearly independent of the density. The one-level system, which has only the ground Landau level filled, has no kinks, the two-level system has one kink at the ground to first level transition and the three-level system has two kinks, one at the ground to first and the other at the first to second-level transition. We have studied each of these systems at three maximum Fermi energies $E_{Fmax} = 2\,m_{e}c^{2}$, $20\,m_{e}c^{2}$ and $200\,m_{e}c^{2}$ and obtained the mass-radius relations of the corresponding stars. 

\item  The most interesting result obtained is that there are possible WDs whose mass exceeds the Chandrasekhar limit and is found to be about $2.3 M_{\odot}$. For instance, for $\epsilon_{Fmax} = 20$ and $B = 2.94 \times 10^{15}$ G, we obtain such a WD with $ \rho_{c} = 2.2 \times 10^{9}$ gm/cc. Interestingly, the nature of mass-radius relations does not depend on the value of $\epsilon_{Fmax}$, however $\epsilon_{Fmax}$ determines how relativistic the system is. For instance, the Chandrasekhar limit is not exceeded for $\epsilon_{Fmax} =2$, no matter what $\rho_{c}$ is.

\item As the magnetic field increases the WDs become more compact in size.

%Out of the equations of state considered in the present work, the system with the lowest magnetic field which gives rise to this mass is the three-level system with $E_{Fmax} = 20\,m_{e}c^{2}$ and $B_{D} = 66.5$ (or $B = 2.94 \times 10^{15}$ G), the corresponding central density being $2.2 \times 10^{9}$ gm/cc. The nature of the mass-radius relations is governed by the fact whether the system is one-level, two-level or three-level and is independent of the value of $E_{Fmax}$, however $E_{Fmax}$ determines how relativistic the system is and so, for instance, the Chandrasekhar mass limit is not exceeded for a low $E_{Fmax}$ (say = $2\,m_{e}c^{2}$) no matter what the central density is. We have also found that as the magnetic field increases the equation of state softens, leading to smaller stars at higher magnetic fields. 

\item The minimum magnetic field required to have a $2.3M_{\odot}$ WD is $B = 2.94 \times 10^{15}$ G. The magnetic field of the original star of radius $R_{\odot}$, which collapses into a WD of radius $\sim 10^{8}$ cm with the above field at its center, then turns out to be $\sim 6 \times 10^{9}$ G, based on the flux freezing theorem. Existence of such stars is not ruled out \cite{shapiro}.

\item In principle, the strong magnetic field causes the pressure to become anisotropic \cite{monica}, \cite{ferrer}. As a result, the combined fluid-magnetic medium develops a magnetic tension \cite{boquet}, leading to a deformation in the WDs along the direction of the magnetic field. The WDs hence could adopt a flattened shape \cite{das}.

\item The flattening effect due to magnetic field leads to super-Chandrasekhar WDs even for smaller magnetic fields. Such relatively weakly magnetized WDs are more probable in nature.

\item We end by addressing the possible reason for not observing such a high field yet in a WD. This could be due to the magnetic screening effects on the surface of the WD if it is an accreting one. In this case the current in the accreting plasma depositing on the surface, presumably creates an induced magnetic moment of sign opposite to that of the original magnetic dipole, thus reducing the surface magnetic field of the WD,  without affecting the central field. Hence by estimating the surface field alone one should not assume the rest.

\end{itemize}

%In this work we have dealt with a constant magnetic field. It would be more realistic to work with a magnetic profile which yields a low field at the surface of the degenerate star and a considerably higher field at the center and solve the problem using the Tolman-Oppenheimer-Volkoff (TOV) equation. However that should not affect the main features of the present work.

\section*{Acknowledgments}

\indent \indent This work was partly supported by the ISRO grant ISRO/RES/2/367/10-11.

\end{document}